\documentclass[iop]{emulateapj}
\usepackage{hyperref}
\usepackage[all]{hypcap} 
\usepackage{amssymb}
\usepackage{color}

\usepackage{graphicx}
\usepackage{float}
\usepackage{amsmath}
\usepackage{txfonts}

\long\def\symbolfootnote[#1]#2{\begingroup%
\def\thefootnote{\fnsymbol{footnote}}\footnote[#1]{#2}\endgroup}

\slugcomment{To appear in the Astronomical Journal}

\shorttitle{Identifying RR Lyrae stars in the SDSS$\times$Pan-STARRS1 Overlapping Area}
\shortauthors{abbas et al.}

\begin{document}
\def\citeapos#1{\citeauthor{#1}'s (\citeyear{#1})}

\title{An optimized Method to Identify RR Lyrae stars in the SDSS$\times$Pan-STARRS1 Overlapping Area Using a Bayesian Generative Technique}

\author{Mohamad Abbas\altaffilmark{1,2}, Eva K. Grebel\altaffilmark{1}, N. F. Martin\altaffilmark{3,4}, N. Kaiser\altaffilmark{5}, W. S. Burgett\altaffilmark{5}, M. E. Huber\altaffilmark{5}, C. Waters\altaffilmark{5}}

\altaffiltext{1}{Astronomisches Rechen-Institut, Zentrum f\"{u}r Astronomie der Universit\"{a}t Heidelberg, M\"{o}nchhofstr. 12--14, D-69120 Heidelberg, Germany}
\altaffiltext{2}{Member of the IMPRS for Astronomy \& Cosmic Physics at the University of Heidelberg and of the Heidelberg Graduate School for Fundamental Physics}
 \email{mabbas@ari.uni-heidelberg.de}
\altaffiltext{3}{Max-Planck-Institut f\"ur Astronomie, K\"onigstuhl 17, D-69117 Heidelberg, Germany}
\altaffiltext{4}{Observatoire astronomique de Strasbourg, Universit\'e de Strasbourg, CNRS, UMR 7550, 11 rue de l'Universit\'e, F-67000 Strasbourg, France}
\altaffiltext{5}{Institute for Astronomy, University of Hawaii at Manoa, Honolulu, HI 96822, USA}

\begin{abstract}

We present a method for selecting RR Lyrae (RRL) stars (or other type of variable stars) in the absence of a large number of multi-epoch data and light curve analyses. Our method uses color and variability selection cuts that are defined by applying a Gaussian Mixture Bayesian Generative Method (GMM) on 636 pre-identified RRL stars instead of applying the commonly used rectangular cuts. Specifically, our method selects 8,115 RRL candidates (heliocentric distances $\textless$ 70 kpc) using GMM color cuts from the Sloan Digital Sky Survey (SDSS) and GMM variability cuts from the Panoramic Survey Telescope and Rapid Response System 1 3$\pi$ survey (PS1). Comparing our method with the Stripe 82 catalog of RRL stars shows that the efficiency and completeness levels of our method are $\sim$77$\%$ and $\sim$52$\%$, respectively. Most contaminants are either non-variable main-sequence stars or stars in eclipsing systems. The method described here efficiently recovers known stellar halo substructures. It is expected that the current completeness and efficiency levels will further improve with the additional PS1 epochs ($\sim$3 epochs per filter) that will be observed before the conclusion of the survey. A comparison between our efficiency and completeness levels using the GMM method to the efficiency and completeness levels using rectangular cuts that are commonly used yielded a significant increase in the efficiency level from $\sim$13$\%$ to $\sim$77$\%$ and an insignificant change in the completeness levels. Hence, we favor using the GMM technique in future studies. Although we develop it over the SDSS$\times$PS1 footprint, the technique presented here would work well on any multi-band, multi-epoch survey for which the number of epochs is limited.

\end{abstract}


\section{Introduction}

Understanding the process of galaxy formation has always been an important goal in astrophysics. In particular, the formation and evolution of disk galaxies still pose many unsolved questions. Many observational studies have focused on the Milky Way as the one disk galaxy that can be studied in the greatest detail (e.g., see the reviews of \citealt{freeman2002,ivezic2012} ). Special emphasis in these studies has been placed on the Galactic stellar halo (e.g., \citealt{johnston2008,schlaufman2009}), the old roughly spherical and extended component of our Galaxy, which is believed to hold important information about the process of galaxy formation.

While accretion of massive systems and in situ star formation processes (e.g., \citealt{yanny2003,juric2008,delucia2008,zolotov2010,font2011, schlaufman2012}) presumably resulted in the formation of the inner halo (Galactocentric radius less than 15 kpc), it is believed that the outer halo formed as a result of accretions and mergers of smaller systems (e.g., \citealt{ibata1995,bullock2001,newberg2003, bullock2005, duffau2006,carollo2007, mccarthy2012,beers2012}). This scenario implies that many of the halo stars were formed in dwarf galaxies outside the Milky Way (e.g., \citealt{bullock2005,abadi2006}). As witnesses of the early phase of the formation of our Galaxy, these halo stars can be used as fossils to trace back the history of our Galaxy (e.g., \citealt{johnston2008,schlaufman2009,zolotov2010}). A complete and a deep map of the halo is vital to find the remnants of the accretion processes (e.g., \citealt{keller2008,bell2008,zolotov2010}). Over the past decade, various halo overdensities and stellar streams have been discovered using different methods and different types of stars. For a summary, see \citet{ivezic2012}. The accreted substructures identified so far mainly seem to consist of old stars. Thus, it is expected that such populations are revealed by maps of RR Lyrae (RRL) stars, these being found only in old stellar populations.

Hence, finding RRL stars and their distances is one way to map the Galactic halo and find its stellar streams. These stars can also be used as objects to study the intrinsic halo population, the distribution, and the gradients in halo metallicity. For instance, the domination of the inner and outer halo by slightly more metal rich and metal poor stars, respectively, and their different global kinematics supports the different scenarios of the formation processes of the inner (in situ formation) and outer (accretion processes) halo \citep{carollo2007,carollo2010}. This evolutionary picture is also supported by studying RRL stars in both parts of the halo (e.g., \citealt{kinman2012}). However, the number of predicted substructures vary substantially (e.g., \citealt{bell2008,deason2011,zinn2013}) and thus more observations are needed. 

Another advantage of using RRL stars to map the Galactic halo is their well-defined mean absolute $V$-band magnitude ($\langle M_{V} \rangle$ = 0.6, \citealt{layden1996}) which can be used to infer their distances in addition to their well studied colors and light curve properties. They are variable horizontal branch stars with periods less than $\sim$ 1 day \citep{smith1995}, so the detection of RRL stars requires repeated observations.

For instance, \citet{watkins2009} and \citet{sesar2010} used data from the Sloan Digital Sky Survey (SDSS; \citealt{fukugita1996,york2000,abazajian2009}) to look for RRL stars in Stripe 82 ($-50\,^{\circ}$ $\textless$ R.A. $\textless$ $59\,^{\circ}$, $-1.25\,^{\circ} \textless$ Dec. $\textless$ $1.25\,^{\circ}$), which was observed around 80 times. The \citet{watkins2009} and \citet{sesar2010} catalogs contain 407 and 483 RRL stars in Stripe 82, respectively, with heliocentric distances ($d_{h}$) in the $\sim$ 4--120 kpc range. According to \citealt{sesar2010}, their catalog has efficiency (fraction of the true RRL stars in the sample) and completeness (fraction of the RRL stars recovered in the sample) levels of $\gtrsim99\%$. Hence, we use the latter catalog as a comparison catalog to compute the efficiency and completeness levels in our study.

Using the SDSS and the Lincoln Near Earth Asteroid Research survey (LINEAR; \citealt{harris1998,sesar2011}), \citet{sesar2013} announced the discovery of $\sim$ 5,000 RRL stars with $d_{h}$ in the 5--30 kpc range that cover $\sim$ 8,000 deg$^2$ of they sky. LINEAR has no spectral filters and has a mean number of 250 observations per object. These RRL stars were selected using SDSS color cuts, LINEAR variability cuts, and light curve analysis.

The Catalina Real-Time Transient Survey (CRTS, \citealt{drake2009,drake2013}) was used to discover $\sim$ 14,000 RRab stars with $d_{h}$ up to 100 kpc using variability statistics, period finding and Fourier fitting techniques \citep{drake2009,drake2013}. Just like LINEAR, CRTS observes the sky repeatedly ($\sim$ 250 times per object) using no spectral filters. 

%\subsection{Our Method}

Using data from the SDSS, Panoramic Survey Telescope and Rapid Response System 1 3$\pi$ survey (hereafter PS1; \citealt{kaiser2002,kaiser2010}), and the CRTS, \citealt{Abbas2014a} (hereafter Paper 1) were able to detect $\sim$ 6,371 RRL stars with an efficiency of $\sim$99$\%$ and $\sim$87$\%$ for RRab and RRc stars, respectively. The high efficiency level obtained was due to the accurate variability statistics and light curve analyses obtained from the CRTS multi-epoch data. The Template Fitting Method (TFM; \citealt{layden1998,layden1999}) and visual inspection were performed on all light curves for a more reliable classification (Paper 1). When light curve analyses are available, the techniques used in Paper 1 can be adopted to detect RRL stars easily. However, light curve analysis is not always possible as not all surveys provide enough multi-epoch data. The technique developed and used in the current paper can be adopted in such surveys with few epochs.

In the current paper, we look for RRL candidates by cross-matching the SDSS data with data from PS1. We show that using a Gaussian Mixture Bayesian Generative Method (GMM, \citealt{AstroML}) to set selection boundary cuts on the SDSS colors and PS1 variability allows one to find RRL stars (or other types of variable stars) even when only a small number of repeated observations are available and light curve analysis is not possible. Our method's efficiency and completeness levels also allow us to detect halo stellar streams and substructures.

A more detailed description of the surveys we used is given in Section \ref{surveys}. In Section \ref{rrlstars}, we study the properties of RRL stars in the SDSS and PS1 photometric systems using more than 600 pre-identified RRL stars. In Section \ref{effcompl}, we describe our method for selecting RRL candidates using the GMM selection boundary cuts for the SDSS colors and the PS1 variability. In the same section, we compute the efficiency and completeness levels of our method by comparing our results with the catalog of RRL stars from \citealt{sesar2010}. Additionally, we compare the efficiency and completeness levels of our GMM method to the efficiency and completeness levels obtained using the rectangular cuts technique. In the same section, we study the properties of the contaminant stars. In Section \ref{rrcandidates}, we apply our color and variability cuts to the whole overlapping footprint between the SDSS and PS1 to find the RRL candidates. In Section \ref{distance}, we derive the distances for our RRL candidates and we use these distances to recover two known halo substructures. The content of the paper is summarized and discussed in Section \ref{D&C}.

\section{Survey Data} \label{surveys}

Our method for searching for RRL stars works by using color and variability information from the SDSS and PS1, respectively. 

\subsection{SDSS And PS1}  \label{ps1}
The SDSS \citep{stoughton2002, abazajian2009} is a deep spectroscopic and photometric survey ($g \textless 23.3$) that uses 5 filters ($u$, $g$, $r$, $i$, and $z$) to survey $\sim$ 12,000 deg$^2$ of the sky. Although most of the SDSS data are based on single-epoch observations, $\sim$ 270 deg$^2$ of the Southern Galactic hemisphere, the so-called Stripe 82, have been observed around 80 times.

%\subsection{PS1}
The PS1 3$\pi$ survey \citep{kaiser2002,kaiser2010} is a $\sim$ 3.5--year (May 2010 -- March 2014) multi-epoch photometric and astrometric survey that is being conducted in Hawaii. The PS1 telescope repeatedly observes the entire sky north of declination 30$^{\circ}$ (3$\pi$ survey). It uses a 1.8 m telescope with a 7 deg$^2$ field of view. It is equipped with the largest digital camera in the world (1.4 Gigapixels). One of its goals is to carry out a photometric and astrometric survey of stars in the Milky Way and the Local Group in 5 bandpasses ($g_{P1}$, $r_{P1}$, $i_{P1}$, $z_{P1}$, and $y_{P1}$) covering the spectral range of 4,000 \AA $<\lambda<$ 10,500 \AA. More information about these filters can be found in \citet{tonry2012}. The PS1 obtains multiple images of three quarters of the celestial sphere in the optical and near-infrared \citep{kaiser2002} to $\sim$ 22 mag in $g_{P1}$ in individual exposures \citep{morganson2012}. Specifically, it is designed to take four exposures per year and area with each of its filters \citep{morganson2012}. By the end of the survey there should be $\sim$ 12 exposures per field and filter. Currently, the average number of observations in each of the five filters is 8 \citep{magnier2013}.

The PS1 was mainly designed to detect potentially hazardous asteroids and near Earth objects (NEOs; \citealt{kaiser2002}). Because it is a deep survey that is repeatedly observing three quarters of the sky, its data are of interest for a wide range of different scientific topics. These topics cover different science areas, from solar system objects to cosmology. The PS1 data are of particular interest also for structural studies of the Milky Way affording a deeper and wider area coverage than previous surveys. When more than $\sim$ 4 epochs are available in at least two filters ($\sim$ 4 epochs in each filter), the repeat observations of the PS1 allow one to identify variable stars such as RRL stars.

%\section{Color and Variability Cuts} \label{selection} 
\section{RRL Stars} \label{rrlstars}

RRL stars are best identified using color cuts, variability cuts, and light curve analysis. Although the colors of RRL stars in the SDSS photometric system have been studied and identified, the lack of variability information and light curve analysis poses difficulties in identifying these stars using the SDSS data alone. The SDSS data are based on single epoch observations with the exception of the overlapping regions and Stripe 82 \citep{sesar2007,bramich2008,sesar2010}. The PS1 is a multi-epoch survey that can be used to study the variability of stars but finding RRL stars using the PS1 data alone is a challenge since the number of repeat observations used in PS1 is small (at most 10 epochs per filter) and the cadence is somewhat irregular. 

Most of the previous studies that looked for RRL stars used a large number of multi-epoch data for each star which allowed them to analyze their light curves. We on the other hand are using the small number of PS1 repeated observations, which makes finding these stars a challenge. Nonetheless, we will demonstrate that using GMM \citep{AstroML} to set selection boundary cuts on the SDSS colors and PS1 variability allows us to find RRL stars to detect halo stellar streams and substructures.

\subsection{The Colors of RRL Stars} \label{rrlcolors} 

The SDSS colors of RRL stars have been studied and characterized using the 483 RRL stars detected in Stripe 82 (\citealt{sesar2010}, and other studies). Since the SDSS $(u-g)$ color serves as a surface gravity indicator for these stars, the range ($\sim$ 0.3 mag) and the root-mean-square ($rms$) scatter ($\sim$ 0.06 mag) are the smallest in this color \citep{ivezic2005}. 

The $g$, $r$, $i$, and $z$ bands from the SDSS are similar to the $g_{P1}$, $r_{P1}$, $i_{P1}$, and $z_{P1}$ bands from the PS1, respectively. However, the $u$ band is used only in the SDSS but not in the PS1, and the $y_{P1}$ band is found only in the PS1 but not in the SDSS. This is due to the difference in the surveys' major scientific goals and in the different sensitivities in the used cameras. The lack of the $u$ filter in the PS1 is a disadvantage when it comes to finding RRL stars.

\begin{figure}
\centering
   \includegraphics[scale=0.42]{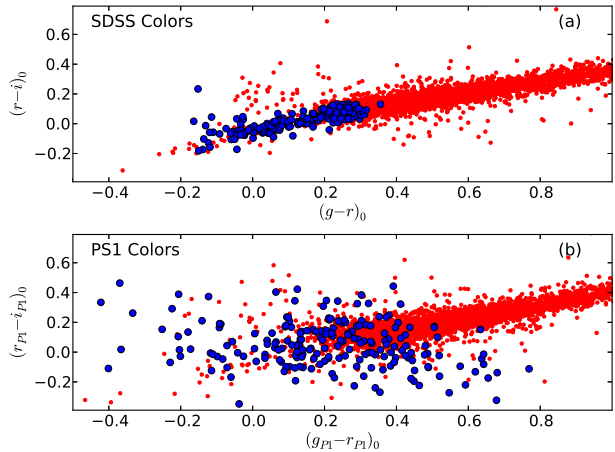}   
  \caption{Illustration of the difference of the colors of RRL stars in the SDSS and PS1 photometric systems. Red dots show a subsample of non-RRL stars in Stripe 82 while blue filled circles show a subsample of the RRL stars detected in the same Stripe \citep{sesar2010}. The scatter of RRL stars in the PS1 plot is due to non-simultaneous $g_{P1}$ and $r_{P1}$ observations by PS1 while the well-defined color region occupied by the RRL stars in the SDSS plot is due to the near-simultaneous imaging observations by the SDSS.}
\label{sdss_ps1_rrcolors}
\end{figure}

Additionally, the SDSS operates in a drift-scanning mode where the sky objects pass through its 5 different filters almost simultaneously. The correct colors of the observed sky objects can then be obtained unless they are variable on very short time scales (i.e., few minutes). Consequently, the SDSS drift-scanning technique gives the correct colors of RRL stars as these stars have periods in the $\sim$ 0.2--1 days range. 

However, the correct colors of RRL stars are not provided with the PS1 photometric system because of the PS1 imaging technique. The PS1 images a selected patch of the sky with different filters at different times. Magnitudes in different filters correspond to different phases for short period variable objects like RRL stars.

Figure \ref{sdss_ps1_rrcolors}a illustrates the $(g-r)$ vs. $(r-i)$ color-color diagram of stars in Stripe 82 from the seventh data release of the SDSS (SDSS DR7; \citealt{abazajian2009}), and Figure \ref{sdss_ps1_rrcolors}b illustrates the PS1 $(g_{P1}-r_{P1})$ vs. $(r_{P1}-i_{P1})$ color-color diagram for the same stars. Red dots represent a subsample of non-RRL stars while blue filled circles represent a subsample of the RRL stars detected in Stripe 82 \citep{sesar2010}. While the RRL stars occupy a small and well-defined region in the SDSS color-color diagram (see Figure \ref{sdss_ps1_rrcolors}a), they are spread out over a large and wide region in the PS1 color-color diagram (see Figure \ref{sdss_ps1_rrcolors}b). This is a result of the different observing techniques used by the SDSS (near-simultaneous imaging using different filters) and PS1 (non-simultaneous imaging).

We base our color cuts for selecting RRL candidates on colors from the SDSS DR7 photometric system and not on the colors from the PS1 photometric system due to the lack of the $u$ band and of the true colors of RRL stars in the latter photometric system.

\begin{figure*}
\centering
   \includegraphics[scale=0.6]{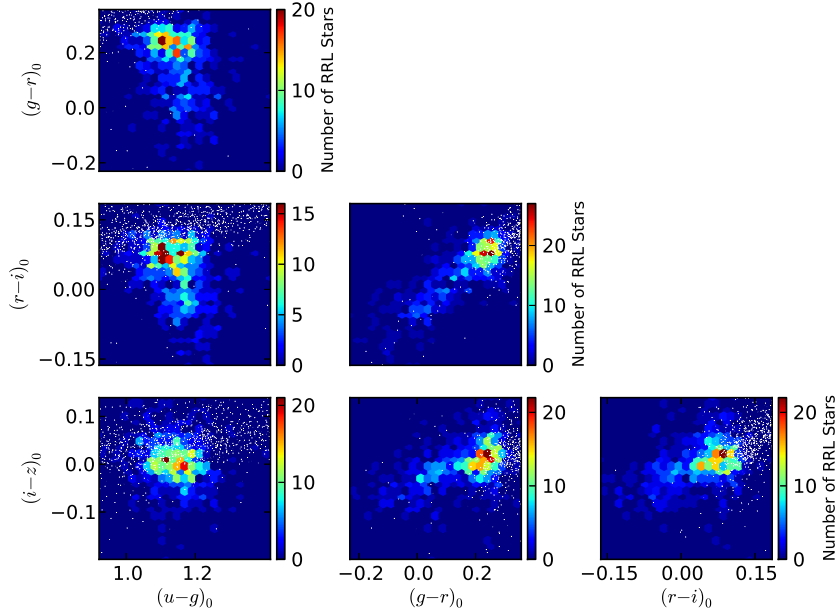}  
  \caption{Different color-color diagrams of the 636 RRL stars in the SDSS photometric system where the red and blue regions reflect large and small numbers of RRL stars, respectively, as indicated by the color bars to the right of each panel. A sample of non-RRL stars are indicated as small white dots to demonstrate the colors of these contaminant stars. These colors are corrected for extinction using the recalibration of \citeapos{schlegel1998} dust map by \citet{schlafly2011}. RRL stars occupy small and well-defined regions in these plots.}
\label{density_sdss_rr_colors}
\end{figure*}

\subsection{Pre-identified Sample of RRL stars} \label{pre-rrl}

We use 636 pre-identified RRL stars selected from the catalogs of RRL stars in the CRTS \citep{drake2013} and LINEAR \citep{sesar2013} surveys for a better characterization of the SDSS colors and PS1 variability properties of RRL stars.

These 636 RRL stars are chosen based on their clean photometry in the SDSS DR7 and PS1 photometric systems. These stars have photometric errors of less than 0.2 in $u$ and less than 0.1 in $g$, $r$, $i$, $z$, $g_{P1}$, and $r_{P1}$. These are primary objects that are not blended or saturated in both surveys and that have been observed more than twice by PS1 in both $g_{P1}$ (N$g_{P1} \ge$ 3) and $r_{P1}$ (N$r_{P1} \ge$ 3), respectively. N$g_{P1}$ and N$r_{P1}$ represent the number of PS1 observations in the $g_{P1}$ and $r_{P1}$ filters, respectively. The two last cuts were applied in order to study the variability of RRL stars in the PS1 multi-epoch data.

We corrected the magnitudes for extinction using the recalibration of \citeapos{schlegel1998} dust map by \citet{schlafly2011}. Since the RRL stars used here are located in areas where the extinction is small (i.e., at high Galactic latitudes), such color corrections can be used. The color densities of the 636 RRL stars in the SDSS photometric system are shown in Figure \ref{density_sdss_rr_colors} where red and blue regions reflect large and small numbers of RRL stars, respectively. A sample of non-RRL stars are also plotted as small white dots to demonstrate the colors of these contaminant stars (i.e., main-sequence stars and stars in eclipsing systems). RRL stars occupy small areas in the color-color diagrams in Figure \ref{density_sdss_rr_colors} and are concentrated in well-defined regions, especially in the $(u-g)$ color, an advantage that helps in finding these stars.

\section{Applying and Testing our Method} \label{effcompl}

It is important to test and maximize the completeness and efficiency levels of our method in selecting RRL stars before we apply our color and variability cuts to the whole area where the SDSS and PS1 data overlap.

For that reason, we define and apply our GMM color and variability boundary cuts to the stars found in Stripe 82. We then compare the Stripe 82 catalog of RRL stars, which has efficiency and completeness levels of $\gtrsim99\%$ \citep{sesar2010} to the RRL stars that our method detects in the same region. RRL stars in \citeapos{sesar2010} catalog span $d_{h}$ between $\sim$ 4 and $\sim$ 120 kpc and $g$ magnitudes between $\sim$ 12.8 and $\sim$ 21.1 mag. There are 374 RRL stars in \citeapos{sesar2010} catalog that are found in the overlapping area covered by PS1 and that are within our magnitude range ($14.0\textless g_{P1} \textless20.0$).

We base our comparison on these 374 RRL stars that are $\gtrsim99\%$ efficient and complete in our magnitude range and sky coverage. We apply all our cuts and selection criteria step by step to stars found in Stripe 82. We then compute the efficiency and completeness levels for each step. 

\subsection{Stripe 82}\label{steps}

%\begin{enumerate} \label{steps}

%\subsubsection{X} \label{steps}
\subsubsection{Step 1}\label{1}
%\item\label{catalog1} 
We start by adopting initial rectangular color cuts from \citet{sesar2010} to avoid downloading all the SDSS DR7 data in Stripe 82 (and later for the whole SDSS$\times$PS1 footprint). Our RRL candidates must first pass the first four initial rectangular color cuts (Equations (6)--(9) in \citealt{sesar2010}):

\begin{equation} \label{1stcolor}
0.75 \textless (u-g) \textless 1.45
\end{equation} 
\begin{equation}
-0.25 \textless (g-r) \textless 0.40
\end{equation} 
\begin{equation}
 -0.20 \textless (r-i) \textless 0.20 
\end{equation} 
\begin{equation} \label{lastcolor}
-0.30 \textless (i-z) \textless 0.30 
\end{equation} 

These are single-epoch color ranges \citep{sesar2010} for RRab and RRc stars corrected for extinction using the \citet{schlegel1998} dust map. The SDSS colors for RRL stars correspond to a random instant in their phase and depend on the time when the near-simultaneous SDSS photometry was obtained. It is thus safe to apply these color criteria to SDSS data but they are not suitable for PS1 data where the color range needs to be larger in order to account for the non-simultaneous observations in the PS1 filters. In order to avoid galaxies, these objects must be flagged as stars ($\tt{type_{SDSS} = 6}$) in the SDSS. They should also be flagged as primary objects ($\tt{mode_{SDSS} = 1}$) with clean photometry in the SDSS DR7 database ($\tt{clean_{SDSS} = 1}$).

Due to the noise and photometric errors resulting from the small number of PS1 epochs that we use in our method, some non-variable sources might appear as variables, especially faint sources with large photometric errors and bright sources that might saturate the CCD camera (see Section \ref{contaminant}). To avoid this, we choose sources that are fainter than 14th and brighter than 20th magnitude in the PS1 $g_{P1}$ filter (Equation (\ref{gp1})). Although PS1 will eventually observe each object around 12 times in each filter, the survey is not finished yet and the average number of detections per star is $\sim$ 8 epochs in each of the PS1 filters. Some of these detections were not taken under good photometric conditions and therefore were flagged as bad sources by the PS1 pipeline. To ensure the reliability of our variability cuts, only clean PS1 detections that are not saturated or blended, and are not flagged as cosmic rays are used in our study \citep{morganson2012}. 

Thus, we only choose stars that have been more than two clean detections in both the $g_{P1}$ and $r_{P1}$ filters (Equations (\ref{ngp1})--(\ref{nrp1})) in order to reliably distinguish variable from non-variable stars:

\begin{equation} \label{gp1}
14.0  \textless g_{P1}  \textless 20.0
\end{equation} 
\begin{equation} \label{ngp1}
N_{g_{P1}} \ge 3
\end{equation} 
\begin{equation} \label{nrp1}
N_{r_{P1}} \ge 3
\end{equation}

Variability cuts in the $i_{P1}$, $z_{P1}$, and $y_{P1}$ filters are applied later. 

In the studied area of Stripe 82, we have $\sim$ 74,000 stars that passed the first four initial SDSS color cuts (Equations (\ref{1stcolor})--(\ref{lastcolor})), the PS1 magnitude cut (Equation (\ref{gp1})), and the PS1 threshold limit of the number of detections in both the $g_{P1}$ and $r_{P1}$ filters (Equations (\ref{ngp1})--(\ref{nrp1})).

Although there are 374 RRL stars in the same area, we missed 85 of them. Around 92$\%$ of these 85 stars did not have more than 2 clean $g_{P1}$ or $r_{P1}$ PS1 detections (Equations (\ref{ngp1})--(\ref{nrp1})) while the rest 8$\%$ of the missed RRL stars did not pass all of the four SDSS color cuts (Equations (\ref{1stcolor})--(\ref{lastcolor})). This leaves us with 289 true RRL stars that we recovered in Stripe 82 (among the $\sim$ 74,000 stars that passed all the conditions in this step). The efficiency and completeness levels are then $\sim$ 0.39$\%$ ($\frac{289}{74,000}$) and 77.3$\%$ ($\frac{289}{374}$), respectively.

 %\item\label{catalog2} 
\subsubsection{Step 2}\label{2}

In order to optimize our color selection of RRL candidates, we define color selection boundaries using the 636 RRL stars (see Section \ref{pre-rrl}) in the SDSS $(u-g)$ vs. $(g-r)$ and $(g-r)$ vs. $(r-i)$ color-color diagrams with GMM \citep{AstroML}. GMM is a Bayesian generative classification method that fits different classes with simple non-correlated Gaussians. These Gaussians are then used to compute the likelihood of a point to belong to each class. The class with the highest likelihood is the predicted result. In our case, GMM uses the colors of the 636 pre-identified RRL stars and compares them to the colors of non-RRL stars to find the GMM color selection boundaries. We choose this method instead of adopting sharp rectangular cuts (e.g., \citealt{vivas2001,sesar2007,sesar2010}) in order to optimize our efficiency and completeness levels when light curve analyses are not possible due to the small number of PS1 observations.

\begin{figure}
\centering
   \includegraphics[scale=0.43]{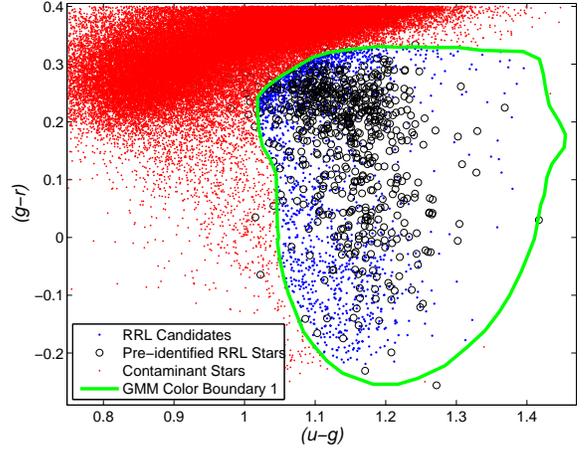}  
  \caption{The $(u-g)$ vs. $(g-r)$ colors of the 636 pre-identified RRL stars used to find the GMM color selection boundary (plotted in green) are shown with black open circles. Stars that fall inside this boundary (blue dots) have RRL-like colors and are retained for further analysis. Stars falling outside the GMM boundary are plotted as red dots and are considered contaminant stars.}
\label{gmm_ugxgr}
\end{figure}

\begin{figure}
\centering
   \includegraphics[scale=0.43]{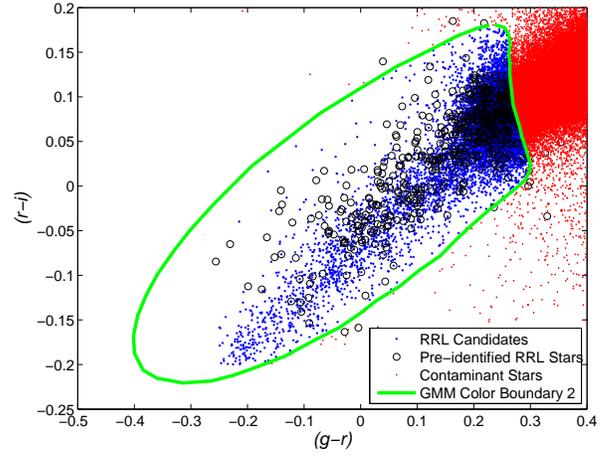}  
  \caption{Same as Figure \ref{gmm_ugxgr}, but showing a $(g-r)$ vs. $(r-i)$ SDSS color-color diagram.}
\label{gmm_grxri}
\end{figure}

In Figure \ref{gmm_ugxgr} and Figure \ref{gmm_grxri}, the GMM color selection boundaries are applied and plotted in green in the $(u-g)$ vs. $(g-r)$ and $(g-r)$ vs. $(r-i)$ color-color diagrams, respectively, for a subsample of stars in Stripe 82. The colors of the 636 pre-identified RRL stars used to find the GMM color selection boundaries are shown with black open circles. Stars that fall inside our GMM selection boundaries are shown as blue dots while stars that fall outside are plotted as red dots. Only stars that fall inside the GMM color selection boundaries in both color-color diagrams ($(u-g)$ vs. $(g-r)$ and $(g-r)$ vs. $(r-i)$) are retained for further analysis.

This step significantly reduces the number of stars in our sample from $\sim$ 74,000 to 1,820 stars, out of which 260 are true RRL stars.

Although the GMM color boundaries are computed using more than 600 well identified RRL stars distributed around the sky, 29 true RRL stars from Step (\ref{1}) did not pass one or both of these GMM color boundary cuts. These stars either have relatively large SDSS magnitude uncertainties that are reflected in their colors or they fall close to, but outside of, our GMM color boundaries.

Because 1,820 stars passed all the cuts in this step (and the cuts in the previous step), and because we were able to recover 260 out of the 374 RRL stars found in Stripe 82, our efficiency level is $\sim$ 14.3$\%$ ($\frac{260}{1,820}$) while the completeness level is 70$\%$ ($\frac{260}{374}$).

\subsubsection{Step 3}\label{3}

%\item\label{catalog3} 

After defining and applying the GMM selection boundaries for the SDSS colors in the previous section, we use the $g_{P1}$, $r_{P1}$, $i_{P1}$, $z_{P1}$, and $y_{P1}$ multi-epoch data from PS1 to distinguish a variable from a non-variable star.

\begin{figure}
\centering
\includegraphics[scale=0.6]{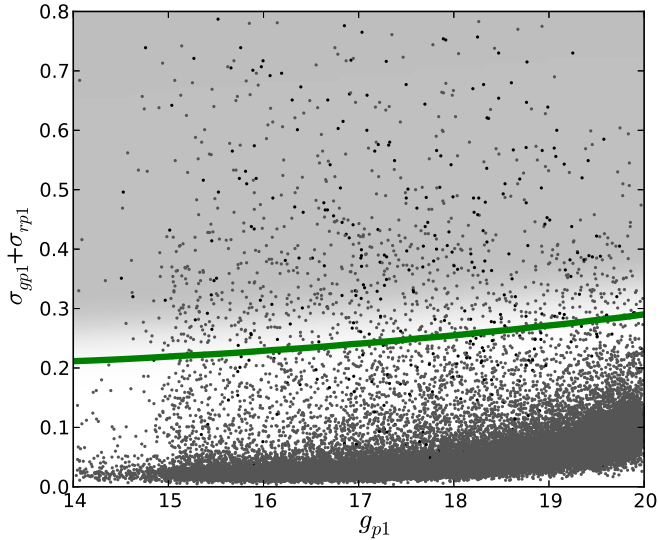}
\caption{The $g_{P1}$ vs. ($\sigma_{g_{P1}}$ + $\sigma_{r_{P1}}$) of a small sample of the stars that passed the SDSS GMM color selection cuts. The green line shows the variability boundary computed by GMM using the 636 pre-identified RRL stars. Stars falling above the variability boundary are retained for further analysis. 
\label{gmm_std}}
\end{figure}

Since we cannot rely on our small number of PS1 detections to phase the light curves and find their periods, we calculate low-order statistics (e.g., standard deviation) and use them to define a GMM selection boundary cut for the $g_{P1}$ magnitudes as a function of the standard deviation in $g_{P1}$ ($\sigma_{g_{P1}}$) plus the standard deviation in $r_{P1}$ ($\sigma_{r_{P1}}$). In Figure \ref{gmm_std}, the GMM variability boundary plotted in green is computed by the GMM method \citep{AstroML} which uses the ($\sigma_{g_{P1}}$ + $\sigma_{r_{P1}}$) values of the 636 pre-identified RRL stars compared to the ($\sigma_{g_{P1}}$ + $\sigma_{r_{P1}}$) values of non-variable stars to find the boundary of the variability cutoff. Although all of these 636 RRL stars are variable sources, only $\sim$ 90$\%$ of them fall above our variability boundary, while $\sim$ 10$\%$ show small or no variability due to the small number of epochs available from PS1. Only stars that fall above our GMM variability boundary are retained for further analysis. These stars have already passed the GMM selection boundaries for the SDSS colors discussed in the previous steps.

In order to be considered as RRL candidates, stars that have more than two clean detections in the $i_{P1}$, $z_{P1}$, and $y_{P1}$ filters must pass the following additional variability criterion: 

\begin{equation} \label{riz_var}
\sigma_{i_{P1}}+\sigma_{z_{P1}}+\sigma_{y_{P1}}\ge 0.1
\end{equation} 

This threshold limit was adopted as more than 90$\%$ of the 636 pre-identified RRL stars (see Section \ref{pre-rrl}) with more than 2 clean detections in the $i_{P1}$, $z_{P1}$, and $y_{P1}$ filters have $\sigma_{i_{P1}}+\sigma_{z_{P1}}+\sigma_{y_{P1}}\ge 0.1$. This criterion is applied to stars with $N_{i_{P1}}\ge 3$, $N_{z_{P1}}\ge 3$, and $N_{y_{P1}}\ge 3$ that have already passed all of our GMM color and variability selection boundaries. Stars that passed our GMM color and variability selection boundaries and that do not have more than two good detections in the $i_{P1}$, $z_{P1}$, and $y_{P1}$ filters are still considered RRL candidates. 

%\end{enumerate}

Applying the GMM variability selection cut (see Figure \ref{gmm_std}) for ($\sigma_{g_{P1}}$ + $\sigma_{r_{P1}}$) and the variability cut in the $i_{P1}$, $z_{P1}$, and $y_{P1}$ filters (see Equation (\ref{riz_var})) to the 1,820 RRL candidates from Step (\ref{2}) reduces the number of RRL candidates to 255 stars, out of which 195 are true RRL stars and 60 are contaminant stars. We discuss the nature of the 60 contaminant stars in Section \ref{contaminant}.

At the same time, 65 RRL stars were lost when moving from Step (\ref{2}) to Step (\ref{3}). These stars did not show a significant amount of variability compared to other variable stars because their number of PS1 epochs is small ($\sim$ 3) and their magnitudes in different detections are not significantly different as they have likely been multiply observed at a relatively close phase.

In this final step, the efficiency significantly increases to $\sim$ 77$\%$ ($\frac{195}{255}$) and the completeness drops to $\sim$ 52$\%$ ($\frac{195}{374}$). This step greatly increases our efficiency level as it gets rid of a large fraction of non-variable stars with colors close to the colors of RRL stars (i.e., main-sequence stars with colors close to the colors of RRL stars).

\subsection{Applying Regular Rectangular Cuts} \label{rectangular_cuts}
We apply the regularly used color and variability rectangular cuts \citep{sesar2010} to the stars in Stripe 82 and compare the results using this technique with the results we achieved using the GMM technique to test weather the latter technique improves the recovery of RRL stars. 

The first step in this technique is similar to Step (\ref{1}) from the previous section where the number of RRL candidates is $\sim$ 74,000 stars of which 289 are known RRL stars. This step requires the SDSS rectangular color cuts, the magnitude cut, and the PS1 threshold limit of the number of detections in both, the $g_{P1}$ and $r_{P1}$ filters. The efficiency and completeness levels are then $\sim$ 0.39$\%$ ($\frac{289}{74,000}$) and 77.3$\%$ ($\frac{289}{374}$), respectively.

Since we are not using the GMM technique in this section, we directly apply straight-line variability cuts in the PS1 filters. Stars with ($\sigma_{g_{P1}}+\sigma_{r_{P1}}\ge 0.22$) that have passed the previous cuts in this section are retained for further analysis. The 636 pre-identified RRL stars were once again used to set the latter cut as more than 90$\%$ of these stars have $\sigma_{g_{P1}}+\sigma_{r_{P1}}\ge 0.22$. Just like in Step (\ref{3}), an additional cut ($\sigma_{i_{P1}}+\sigma_{z_{P1}}+\sigma_{y_{P1}}\ge 0.1$) is applied for the retained stars with $N_{i_{P1}}\ge 3$, $N_{z_{P1}}\ge 3$, and $N_{y_{P1}}\ge 3$. Retained stars that do not have more than two good detections in the $i_{P1}$, $z_{P1}$, and $y_{P1}$ filters are still considered RRL candidates.

There are $\sim$ 1,600 stars that passed all of our cuts in this section of which 205 are known RRL stars. This yields efficiency and completeness levels of $\sim$ 13$\%$ ($\frac{205}{1,600}$) and $\sim$ 54$\%$ ($\frac{205}{374}$), respectively.

\begin{figure}
\centering
\includegraphics[scale=0.47]{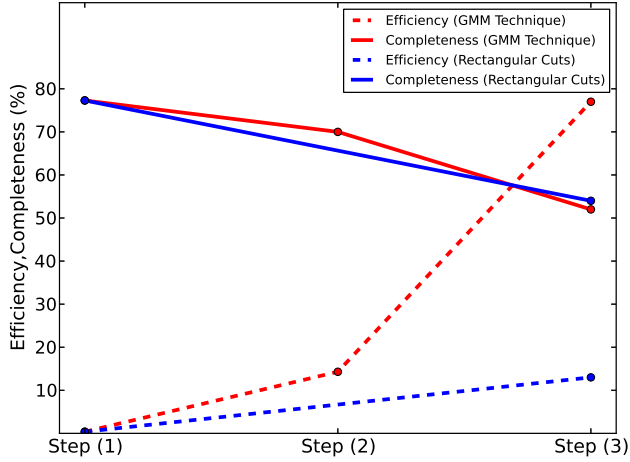}
\caption{A comparison between the efficiency (dashed lines) and completeness (solid lines) levels on each step resulting from the GMM (in red) and rectangular (in blue) cuts techniques. The dependence of our GMM completeness and efficiency levels on Step (\ref{1}): the magnitude, initial color, and number of detection cuts (Equations (\ref{1stcolor})--(\ref{nrp1})), Step (\ref{2}): the SDSS GMM color boundary cuts, and Step (\ref{3}): the GMM variability boundary cut are also shown. 
\label{eff_compl}}
\end{figure}

The dependencies of the efficiency (dashed lines) and completeness (solid lines) levels in each step resulting from the GMM and rectangular cut techniques are plotted with red and blue lines in Figure \ref{eff_compl}, respectively. Although there was no significant change in the completeness level when using the rectangular cuts compared to the GMM technique, the efficiency level increased from $\sim$ 13$\%$ ($\frac{205}{1,600}$, using rectangular cuts) to 77$\%$ ($\frac{195}{255}$, using the GMM technique). Hence, we favor using the GMM technique in future studies.

\subsection{Contaminant Stars} \label{contaminant}
To understand the nature of the contaminant stars, we look for multi-epoch data in the CRTS database for the 60 contaminant stars we found in Stripe 82. 56 out of the 60 contaminant stars are found in the CRTS database and have been observed between $\sim$ 40 and 500 times.

Almost 40$\%$ of these stars showed no variability using the multi-epoch data from CRTS, which makes them non-variable stars that have passed our variability cuts. These stars were observed only three to four times with PS1 and have magnitudes close to our bright ($g_{P1} \sim 14.0$ mag) and faint ($g_{P1} \sim 20.0$ mag) magnitude cuts. Hence, it is not surprising that some non-variable sources passed our variability cuts as their variability statistics are based on a small number of observations where a single noisy epoch can bias the statistics and make a non-variable source appear as a variable one, and vice versa. 

%and SX Phe (P = 0.114 days) 
%, and (c): SX Phe stars (ID$\_{CRTS}$ = 1109088010161

\begin{figure}
\centering
\includegraphics[scale=0.47]{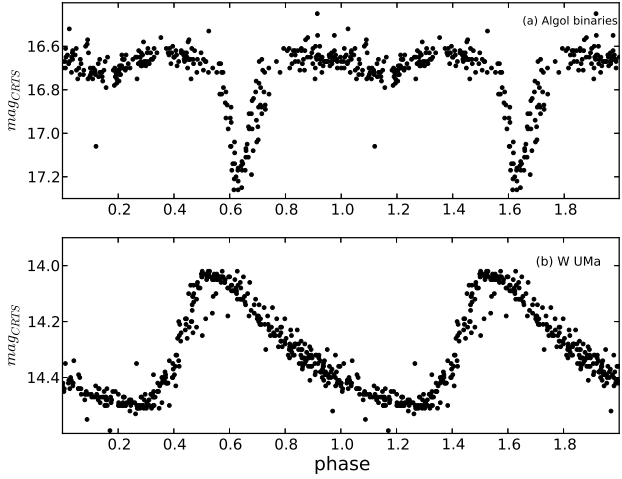}
\caption{Phased light curves of (a): Stars in Algol binary systems ($\tt ID_{CRTS}$ = 18720940), (b): W UMa stars ($\tt ID_{CRTS}$ = 1109081021295).
\label{conta_2lcs}}
\end{figure}

The remaining 60$\%$ of the contaminant stars in Stripe 82 appeared as non-RRL variable stars using the CRTS database. Their variability statistics reflected a change in their brightness over time but the shape of their light curve showed that most of them are W Ursae Majoris (W UMa), Algol binaries, $\delta$ Scuti, and SX Phe stars \citep{palaversa2013}. Samples of the phased light curves for Algol binaries (P = 0.6684 days) and W Uma (P = 0.3 days) stars that are contaminating our RRL stars are shown in panels (a) and (b) of Figure \ref{conta_2lcs}, respectively. We were able to recover the correct type and periods of these stars using the CRTS multi-epoch data. Using the current PS1 data available, there is no way of getting rid of all the contaminants.

\begin{figure}
\centering
\includegraphics[scale=0.47]{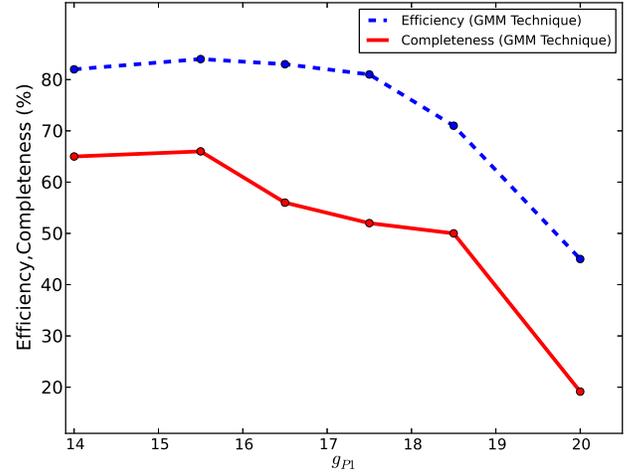}
\caption{The decrease in the efficiency (blue dashed line) and completeness (red solid line) levels as a function of magnitude reflects the increase in contamination for fainter stars. 
\label{MagBin_eff_compl}}
\end{figure}

With the $\sim$ 77$\%$ ($\frac{195}{255}$) efficiency level computed in the previous section, we know that $\sim$ 23$\%$ of our RRL candidates are non-RRL stars (mainly non-variable stars and stars in eclipsing systems). However, we show in Section \ref{streams} that we are still able to detect halo substructures with such a contamination level. Additionally, the efficiency and completeness levels will be improved when more PS1 epochs are available in the near future. Our method can be useful in detecting RRL stars in surveys other than the PS1 where the number of detections per star is also small. Our efficiency and completeness levels as a function of $g_{P1}$ magnitudes are plotted in blue and red lines in Figure \ref{MagBin_eff_compl}, respectively. The decrease in the efficiency and completeness levels as a function of magnitude reflects the increase in contamination for fainter stars.

\section{RRL Candidates} \label{rrcandidates}
Knowing that our efficiency and completeness levels are 77$\%$ ($\frac{195}{255}$) and $\sim$ 52$\%$ ($\frac{195}{374}$), respectively, we apply our method to the whole SDSS$\times$PS1 overlapping footprint.

In the mentioned area, around 130,000 stars passed the first four initial SDSS color cuts (Equations (\ref{1stcolor})--(\ref{lastcolor})), the PS1 magnitude cut (Equation (\ref{gp1})), and the minimum number of PS1 epoch cuts (Equations (\ref{ngp1})--(\ref{nrp1})). These stars have also passed the two GMM selection boundaries in the SDSS colors defined and applied in Step (\ref{2}) of Section \ref{steps}.

\begin{figure}
\centering
\includegraphics[scale=0.47]{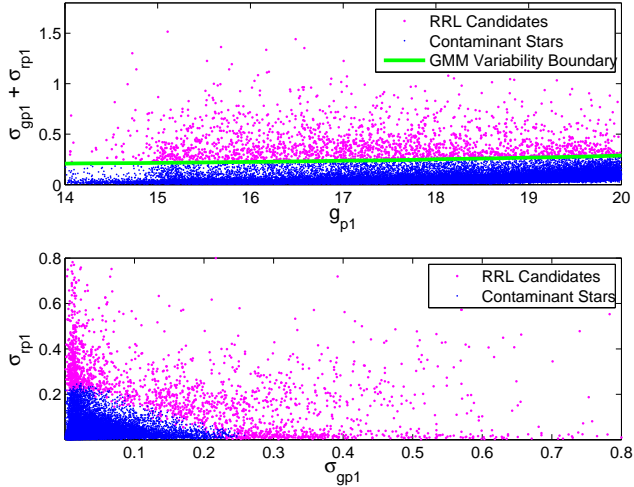}
\caption{The upper panel illustrates how we apply our GMM variability selection boundary (green line) cut to distinguish variable (magenta dots) from non-variable (blue dots) stars in a $g_{P1}$ vs ($\sigma_{g_{P1}}$ + $\sigma_{r_{P1}}$) plot. Stars falling above our GMM variability boundary are considered to be RRL candidates. The lower panel shows the distribution of $\sigma_{g_{P1}}$ vs. $ \sigma_{r_{P1}}$ of the same stars plotted in the upper panel.
\label{GMM_stdgr}}
\end{figure}

Finally, we apply the GMM variability selection cut from Step (\ref{3}) of Section \ref{steps} to these 130,000 stars. To illustrate this, we plot the $g_{P1}$ vs ($\sigma_{g_{P1}}$ + $\sigma_{r_{P1}}$) distribution for the sample of stars (spanning $\sim$ 100 deg$^2$ of the sky) that passed our GMM color boundaries in the upper panel of Figure \ref{GMM_stdgr}. Stars falling below our GMM variability boundary (green line) are plotted as blue dots and are considered non-variable stars. Stars passing the boundary are plotted as magenta dots and are considered RRL candidates. The lower panel of Figure \ref{GMM_stdgr} illustrates the distribution of the same stars, but showing a $\sigma_{g_{P1}}$ vs $\sigma_{r_{P1}}$ plot. 

An additional variability cut was applied to all of our RRL candidates with $N_{i_{P1}}\ge 3$, $N_{z_{P1}}\ge3$, and $N_{y_{P1}}\ge 3$. These stars must pass the $i_{P1}$, $z_{P1}$, and $y_{P1}$ variability cut defined in Equation (\ref{riz_var}) ($\sigma_{i_{P1}}+\sigma_{z_{P1}}+\sigma_{y_{P1}}\ge 0.1$). Stars that passed all of our previous cuts and that do not have more than two good detections in the $i_{P1}$, $z_{P1}$, and $y_{P1}$ filters are still considered RRL candidates.

Only 6$\%$ of the 130,000 stars passed these variability cuts which leaves us with 8,115 RRL candidates. Based on the analysis in Section \ref{contaminant}, we believe that $\sim$ 23$\%$ of these RRL candidates are non-RRL stars (mainly non-variable stars and stars in eclipsing systems).

%The histograms of the distribution of the number of clean detections of our RRL candidates in the $g_{P1}$ (in green) and $r_{P1}$ (in red) filters are shown in the upper and lower panel of Figure \ref{hist_det}, respectively, where the average number of clean detections is $\sim$ 6 in both filters. 

\section{Distances Of RRL Stars} \label{distance}

One of the advantages of RRL stars is their well defined mean absolute $\langle V \rangle$ magnitude which makes it straightforward to find estimates for their distances.

\citet{ivezic2008} calculated the mean halo metallically and obtained [Fe/H] = $-1.5$ dex with $rms_{[Fe/H]}$ $\sim$ 0.32 dex. The mean halo metallically of [Fe/H] $\sim$ $-1.5$ dex has been also used and confirmed (e.g., \citealt{vivas2006,sesar2010,zinn2013}) in different studies including \citeapos{kollmeier2013} recent study of RRc stars by statistical parallax.

Thus, we adopt RRL star mean halo metallicity of $-1.5$ dex and use Equation (\ref{MvFe}) \citep{cacciari2003} to calculate the mean absolute magnitude of RRL stars:

\begin{equation} \label{MvFe}
M_{V} = (0.23 \pm 0.04)\mathrm{[Fe/H]}+(0.93 \pm 0.12)
\end{equation} 

Adopting [Fe/H] = $-1.5$ $\pm$ 0.32 dex introduces $rms_{M_v}$ of $\sim$ 0.1 mag. The $\langle V \rangle$ magnitudes are calculated using Equation (\ref{VMag}), which was adopted from \citealt{ivezic2005}:

\begin{equation} \label{VMag}
\langle V \rangle= r - 2.06(g-r) + 0.355
\end{equation} 

where the $g$ and $r$ SDSS measurements have been corrected for interstellar reddening \citep{schlegel1998,schlafly2011}. Equation (\ref{VMag}) corrects biases that come from the single SDSS epochs for RRL stars that were taken at unknown phases and computes $\langle V \rangle$ with $rms_{\langle V \rangle}$ $\sim$ 0.12 mag \citep{ivezic2008}.

Finally, using Equation (\ref{Heli_Distx}), the heliocentric distance ($d_{h}$, in parsecs), is determined with a $\sim$ 7$\%$ fractional error after taking all the mentioned sources of uncertainties into account:

\begin{equation} \label{Heli_Distx}
d_{h} = 10^{(\langle V \rangle-M_{V}+5)/5}
\end{equation} 

Our 8,115 RRL candidates have $d_{h}$ in the $\sim$ 3--70 kpc distance range.

\subsection{Halo Structure}  \label{streams}

\begin{figure}
\centering
\includegraphics[scale=0.47]{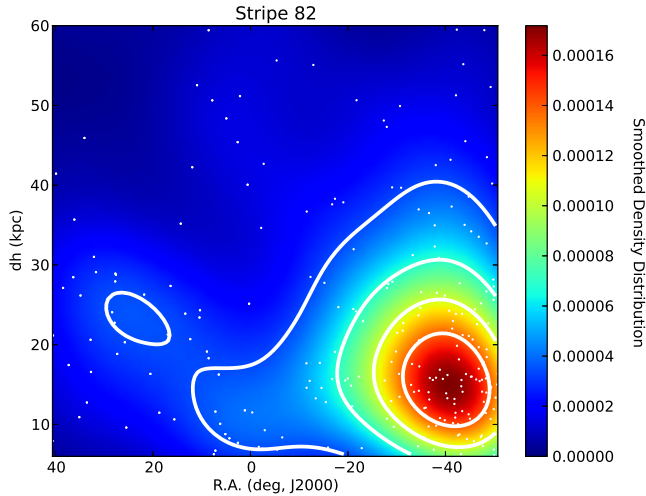}
\caption{The number density distribution of the RRL stars in Stripe 82 is plotted with scaled density levels that are accentuated by the white contours. The Hercules\--Aquila cloud appears at R.A. $\sim$ $-40\,^{\circ}$ and $d_{h}$ between $\sim$ 8 and $\sim$ 24 kpc while the Sagittarius dSph tidal stream is detected at R.A. $\sim$ $30\,^{\circ}$ and $d_{h}$ $\sim$ 23 kpc. Negative values of R.A. were used for better visualization only (R.A. = R.A. + 360$^{\circ}$ when R.A. $\textless$ 0$^{\circ}$)
\label{S82_south}}
\end{figure}

Using the 255 RRL candidates we detected in Stripe 82, we look for halo substructures in our covered distance range. We plot the number density distribution of these 255 RRL candidates in Figure \ref{S82_south}. This plot includes our 60 contaminant stars in Stripe 82 (if we assume that \citeapos{sesar2010} catalog of RRL stars is $\gtrsim99\%$ complete). 

The density of the points that is accentuated by the white contours is shown in scaled density levels. The smoothed surface regions with a high number of stars are indicated in red while regions with low number of stars are indicated in dark blue. We recover the Hercules\--Aquila cloud \citep{belokurov2007} at R.A.\footnote{Add 360\,$^{\circ}$ to obtain the correct values of R.A. when R.A. $\textless$ 0$^{\circ}$. Negative values of R.A. were used for better visualization only.} $\sim$ $-40\,^{\circ}$ and $d_{h}$ between $\sim$ 8 and $\sim$ 24 kpc. The trailing arm of the Sagittarius dwarf spheroidal's (dSph) tidal stream \citep{majewski2003,law2010} is also recovered at R.A. $\sim$ $30\,^{\circ}$ and $d_{h}$ $\sim$ 23 kpc.

Both of our recovered substructures were seen using the $\sim$ 99$\%$ complete and efficient catalog of RRL stars in Stripe 82 \citep{sesar2010}. Although our method is not as efficient and complete as the mentioned catalog, Figure \ref{S82_south} proves that the efficiency and completeness levels we achieved are good enough to select RRL candidates to trace stellar streams and substructures in spite of the inclusion of contaminant stars. Stripe 82 was visited $\sim$ 80 times by the SDSS, which made it relatively easy to find its RRL stars using light curve analysis \citep{sesar2010}. In contrast, it was more difficult to find RRL stars in our study using only the SDSS colors and PS1 variability because of the small number of multi-epoch data available from PS1. 

Nevertheless, we recovered $\sim$ 52$\%$ of the RRL stars ($d_{h}\textless 70$ kpc) not only in the Stripe 82 region, but in the whole SDSS$\times$PS1 overlapping footprint. A detailed analysis of the distribution of the identified RRL candidates will be presented in a future paper.

Having additional PS1 epochs will improve the quality of our variability statistics which will improve the separation between variable and non-variable stars. Using the CRTS data, we showed in Section \ref{contaminant} that 40$\%$ of our contaminant stars are non-variable sources and have small number of PS1 epochs. We expect to get rid of at least 60$\%$ of these non-variable contaminant stars when more PS1 epochs ($\sim$ 15 epochs in all filters) are available. However, it will not be possible to get rid of all of the contaminant stars as the number of PS1 epochs will not be sufficient to distinguish RRL from non-RRL variable stars using light curve and period analysis. Furthermore, having additional PS1 epochs will improve our completeness level as we missed many RRL stars due to the PS1 threshold limit of the number of detections in both the $g_{P1}$ and $r_{P1}$ filters (Equations (\ref{ngp1})--(\ref{nrp1})). We expect the efficiency and completeness level to increase to at least $\sim$ 83$\%$ and $\sim$ 65$\%$ when all of the PS1 epochs are available. Having additional epochs will also allow us to study stars in the halo that are further than 70 kpc ($d_{h}$).

\section{Summary} \label{D&C}

In this study, we combine data from two different sky surveys (SDSS and PS1) to look for RRL candidates in the halo. We select the RRL candidates using SDSS color cuts and PS1 variability cuts. We show that using a GMM method to define GMM boundary cuts optimizes the efficiency and completeness levels to select RRL stars (or other type of variable stars) when light curve analyses are not available.

We start by adopting initial color cuts for RRL stars from \citet{sesar2010}. In order to optimize the selection of our RRL candidates, we use 636 pre-identified RRL stars from CRTS and LINEAR to define GMM color selection boundaries in the SDSS $(u-g)$ vs. $(g-r)$ and $(g-r)$ vs. $(r-i)$ color-color diagrams in addition to a GMM variability boundary cut for the ($g_{P1}$ vs. $\sigma_{g_{P1}}$ $+$ $\sigma_{r_{P1}}$) diagram. We applied another variability cut in the $i_{P1}$, $z_{P1}$, and $y_{P1}$ filters from PS1. A comparison between our efficiency and completeness levels using the GMM method to the efficiency and completeness levels using rectangular cuts that are commonly used yielded a significant increase in the efficiency level from $\sim$ 13$\%$ ($\frac{205}{1,600}$) to $\sim$ 77$\%$ ($\frac{195}{255}$) and an insignificant change in the completeness levels. Hence, we favor using the GMM technique in future studies.

We used the multi-epoch data from the CRTS database to study the properties of our contaminant stars found in Stripe 82. Around 40$\%$ of the contaminant stars showed no sign of variability in the CRTS data. Because these stars have between $\sim$ 40 and 500 CRTS epochs compared to $\sim$ 8 epochs in $g_{P1}$ and $r_{P1}$, we favor the CRTS variability statistics and consider that these stars are contaminating our RRL candidates sample. Noisy detections, poor seeing, and non photometric conditions in the PS1 filters are the reasons for why these stars appeared to be variables in the latter photometric system. Although the remaining 60$\%$ of the contaminant stars in Stripe 82 showed variability using the CRTS data, their variability properties indicate that most of them are W UMa, Algol binaries, $\delta$ Scuti, and SX Phe stars.

Having achieved our best efficiency (77$\%$) and completeness (52$\%$) levels, we apply our selection criteria and cuts to the whole SDSS$\times$PS1 overlapping footprint. Our technique yielded the detection of 8,115 RRL candidates. From the analysis in Section \ref{contaminant}, we believe that $\sim$ 23$\%$ of our RRL candidates are non-RRL stars (mainly non-variable stars and stars in eclipsing systems). Since light curve analysis is not possible in our study, we believe that achieving such a high efficiency and small contamination level reflects the success of our method. With the current PS1 data available, there is no way of getting rid of the contaminants. But it is plausible to assume that getting the remaining PS1 epochs yet to be observed ($\sim$ 3 epochs per filter) would eliminate more contaminant stars and recover more RRL stars. Our method can be applied to data from any multi-band survey where the number of multi-epoch data is small.

We obtain distance estimates for our RRL stars to test if we are still able to detect halo stellar streams and substructures with our efficiency and completeness levels. Although $\sim$ 23$\%$ of the 255 RRL candidates in Stripe 82 are not true RRL stars and although we missed $\sim$ 50$\%$ of the known RRL stars within the magnitude range considered here, we were still able to recover the Hercules\--Aquila cloud and the arm of the Sagittarius dSph tidal stream (see Figure \ref{S82_south}). This proves that our method is good enough to detect some of the halo substructures and stellar streams in the halo.

The technique developed in this paper can be adopted to optimize the selection of a specific type of variable stars when light curve analyses are not possible while the technique developed in Paper 1 can be adopted when a large number of repeated observations are available. We used both techniques to find RRL stars in the halo that we will use in a forthcoming paper to present a more detailed map of halo substructure.

\acknowledgments
%Acknowledgments

We thank the referee for comments and constructive suggestions that helped to improve the manuscript. We thank B. Sesar for helpful discussion that improved the quality of this paper. M.A., E.K.G., and N.F.M acknowledge support by the Collaborative Research Center ``The Milky Way System" (SFB 881, subproject A3) of the German Research Foundation (DFG). N.F.M. gratefully acknowledges the CNRS for support through PICS project PICS06183. 

The Pan-STARRS1 Surveys (PS1) have been made possible through contributions of the Institute for Astronomy, the University of Hawaii, the Pan-STARRS Project Office, the Max-Planck Society and its participating institutes, the Max Planck Institute for Astronomy, Heidelberg and the Max Planck Institute for Extraterrestrial Physics, Garching, The Johns Hopkins University, Durham University, the University of Edinburgh, Queen's University Belfast, the Harvard-Smithsonian Center for Astrophysics, the Las Cumbres Observatory Global Telescope Network Incorporated, the National Central University of Taiwan, the Space Telescope Science Institute, the National Aeronautics and Space Administration under Grant No. NNX08AR22G issued through the Planetary Science Division of the NASA Science Mission Directorate, the National Science Foundation under Grant No. AST-1238877, the University of Maryland, and Eotvos Lorand University (ELTE). 

Funding for the SDSS and SDSS-II has been provided by the Alfred P. Sloan Foundation, the Participating Institutions, the National Science Foundation, the U.S. Department of Energy, the National Aeronautics and Space Administration, the Japanese Monbukagakusho, the Max Planck Society, and the Higher Education Funding Council for England. The SDSS Web Site is http://www.sdss.org/. The SDSS is managed by the Astrophysical Research Consortium for the Participating Institutions. The Participating Institutions are the American Museum of Natural History, Astrophysical Institute Potsdam, University of Basel, University of Cambridge, Case Western Reserve University, University of Chicago, Drexel University, Fermilab, the Institute for Advanced Study, the Japan Participation Group, Johns Hopkins University, the Joint Institute for Nuclear Astrophysics, the Kavli Institute for Particle Astrophysics and Cosmology, the Korean Scientist Group, the Chinese Academy of Sciences (LAMOST), Los Alamos National Laboratory, the Max-Planck-Institute for Astronomy (MPIA), the Max-Planck-Institute for Astrophysics (MPA), New Mexico State University, Ohio State University, University of Pittsburgh, University of Portsmouth, Princeton University, the United States Naval Observatory, and the University of Washington.

The CRTS is supported by the U.S. National Science Foundation under grants AST-0909182 and CNS-0540369. The work at Caltech was supported in part by the NASA Fermi grant 08-FERMI08-0025 and by the Ajax Foundation. The CSS survey is funded by the National Aeronautics and Space Administration under grant No. NNG05GF22G issued through the Science Mission Directorate Near-Earth Objects Observations Program.


\begin{thebibliography}{}



%A

\bibitem[Abbas et al.(2014)]{Abbas2014a} Abbas, M.~A., Grebel, 
E.~K., Martin, N.~F., et al.\ 2014, arXiv:1404.4823 

\bibitem[Abadi et al.(2006)]{abadi2006} Abadi, M.~G., Navarro, 
J.~F., \& Steinmetz, M.\ 2006, \mnras, 365, 747 
\bibitem[Abazajian et al.(2009)]{abazajian2009} Abazajian, K.~N., 
Adelman-McCarthy, J.~K., Ag{\"u}eros, M.~A., et al.\ 2009, \apjs, 182, 543 

%B
\bibitem[Beers et al.(2012)]{beers2012} Beers, T.~C., Carollo, 
D., Ivezi{\'c}, {\v Z}., et al.\ 2012, \apj, 746, 34 
\bibitem[Bell et al.(2008)]{bell2008} Bell, E.~F., Zucker, 
D.~B., Belokurov, V., et al.\ 2008, \apj, 680, 295 
%\bibitem[Belokurov et al.(2006)]{belokurov2006} Belokurov, V., Zucker, D.~B., Evans, N.~W., et al.\ 2006, \apjl, 642, L137 
\bibitem[Belokurov et al.(2007)]{belokurov2007} Belokurov, V., Evans, 
N.~W., Bell, E.~F., et al.\ 2007, \apjl, 657, L89 
%\bibitem[Bernard et al.(2008)]{bernard2008} Bernard, E.~J., Gallart, C., Monelli, M., et al.\ 2008, \apjl, 678, L21 
\bibitem[Bullock et al.(2001)]{bullock2001} Bullock, J.~S., 
Kravtsov, A.~V., \& Weinberg, D.~H.\ 2001, \apj, 548, 33 
\bibitem[Bullock 
\& Johnston(2005)]{bullock2005} Bullock, J.~S., \& Johnston, K.~V.\ 2005, \apj, 635, 931 

\bibitem[Bramich et al.(2008)]{bramich2008} Bramich, D.~M., Vidrih, 
S., Wyrzykowski, L., et al.\ 2008, \mnras, 386, 887 


%C
\bibitem[Cacciari 
\& Clementini(2003)]{cacciari2003} Cacciari, C., \& Clementini, G.\ 2003, Stellar Candles for the Extragalactic Distance Scale (Lecture Notes in Physics), Vol. 635, ed. D. Alloin \& W. Gieren (Berlin: Springer), 105

\bibitem[Carollo et al.(2007)]{carollo2007} Carollo, D., Beers, 
T.~C., Lee, Y.~S., et al.\ 2007, \nat, 450, 1020 
\bibitem[Carollo et al.(2010)]{carollo2010} Carollo, D., Beers, 
T.~C., Chiba, M., et al.\ 2010, \apj, 712, 692 

%bibitem[Cooper et al.(2010)]{cooper2010} Cooper, A.~P., Cole, S., Frenk, C.~S., et al.\ 2010, \mnras, 406, 744 


%D
\bibitem[Deason et al.(2011)]{deason2011} Deason, A.~J., 
Belokurov, V., \& Evans, N.~W.\ 2011, \mnras, 416, 2903 

\bibitem[De Lucia 
\& Helmi(2008)]{delucia2008} De Lucia, G., \& Helmi, A.\ 2008, \mnras, 391, 14 

\bibitem[Duffau et al.(2006)]{duffau2006} Duffau, S., Zinn, R., Vivas, A.~K., et al.\ 2006, \apjl, 636, L97 

\bibitem[Drake et al.(2009)]{drake2009} Drake, A.~J., Djorgovski, 
S.~G., Mahabal, A., et al.\ 2009, \apj, 696, 870 
\bibitem[Drake et al.(2013)]{drake2013} Drake, A.~J., Catelan, 
M., Djorgovski, S.~G., et al.\ 2013, \apj, 763, 32 

%E

%F
\bibitem[Font et al.(2011)]{font2011} Font, A.~S., McCarthy, 
I.~G., Crain, R.~A., et al.\ 2011, \mnras, 416, 2802 

\bibitem[Fukugita et al.(1996)]{fukugita1996} Fukugita, M., Ichikawa, T., Gunn, J.~E., et al.\ 1996, \aj, 111, 1748 



\bibitem[Freeman 
\& Bland-Hawthorn(2002)]{freeman2002} Freeman, K., \& Bland-Hawthorn, J.\ 2002, \araa, 40, 487 


%G
%\bibitem[Grillmair(2006)]{grillmair2006} Grillmair, C.~J.\ 2006, \apjl, 645, L37 

%H
\bibitem[Harris(1998)]{harris1998} Harris, A.~W.\ 1998, \planss, 46, 283
%\bibitem[Hattori et al.(2013)]{hattori2013} Hattori, K., Yoshii, Y., Beers, T.~C., Carollo, D., \& Lee, Y.~S.\ 2013, \apjl, 763, L17  
%\bibitem[Haschke et al.(2012)]{haschke2012a} Haschke, R., Grebel, E.~K., Frebel, A., et al.\ 2012, \aj, 144, 88 
%\bibitem[Helmi \& White(1999)]{helmi1999} Helmi, A., \& White, S.~D.~M.\ 1999, \mnras, 307, 495 
% \bibitem[Helmi(2008)]{helmi2008} Helmi, A.\ 2008, \aapr, 15, 145 

%I
%\bibitem[Ibata et al.(1995)]{ibata1995} Ibata, R.~A., Gilmore, G., \& Irwin, M.~J.\ 1995, \mnras, 277, 781 
%\bibitem[Ibata et al.(2001)]{ibata2001} Ibata, R., Irwin, M., Lewis, G.~F., \& Stolte, A.\ 2001, \apjl, 547, L133 
%\bibitem[Ivezi{\'c} et al.(2000)]{ivezic2000} Ivezi{\'c}, {\v Z}., Goldston, J., Finlator, K., et al.\ 2000, \aj, 120, 963

\bibitem[Ibata et al.(1995)]{ibata1995} Ibata, R.~A., Gilmore, 
G., \& Irwin, M.~J.\ 1995, \mnras, 277, 781 
 
\bibitem[Ivezi{\'c} et al.(2005)]{ivezic2005} Ivezi{\'c}, {\v Z}., 
Vivas, A.~K., Lupton, R.~H., \& Zinn, R.\ 2005, \aj, 129, 1096 
\bibitem[Ivezi{\'c} et al.(2008)]{ivezic2008} Ivezi{\'c}, {\v Z}., 
Sesar, B., Juri{\'c}, M., et al.\ 2008, \apj, 684, 287 

\bibitem[Ivezi{\'c} et 
al.(2012)]{ivezic2012} Ivezi{\'c}, {\v Z}., Beers, T.~C., \& Juri{\'c}, M.\ 2012, \araa, 50, 251 

%J
%\bibitem[Johnston(1998)]{johnston1998} Johnston, K.~V.\ 1998, \apj, 495, 297 
\bibitem[Johnston et al.(2008)]{johnston2008} Johnston, K.~V., 
Bullock, J.~S., Sharma, S., et al.\ 2008, \apj, 689, 936 
\bibitem[Juri{\'c} et al.(2008)]{juric2008} Juri{\'c}, M., 
Ivezi{\'c}, {\v Z}., Brooks, A., et al.\ 2008, \apj, 673, 864 

%K
\bibitem[Kaiser et al.(2002)]{kaiser2002} Kaiser, N., Aussel, H., 
Burke, B.~E., et al.\ 2002, \procspie, 4836, 154 
\bibitem[Kaiser et al.(2010)]{kaiser2010} Kaiser, N., Burgett, W., 
Chambers, K., et al.\ 2010, \procspie, 7733, 12
\bibitem[Keller et al.(2008)]{keller2008} Keller, S.~C., Murphy, 
S., Prior, S., Da Costa, G., \& Schmidt, B.\ 2008, \apj, 678, 851 
\bibitem[Kollmeier et al.(2013)]{kollmeier2013} Kollmeier, J.~A., 
Szczygie{\l}, D.~M., Burns, C.~R., et al.\ 2013, \apj, 775, 57 
%\bibitem[Koposov et al.(2012)]{koposov2012} Koposov, S.~E.,  Belokurov, V., Evans, N.~W., et al.\ 2012, \apj, 750, 80 
\bibitem[Kinman et al.(2012)]{kinman2012} Kinman, T.~D., Cacciari, 
C., Bragaglia, A., Smart, R., \& Spagna, A.\ 2012, \mnras, 422, 2116 


%\bibitem[Kepley et al.(2007)]{kepley2007} Kepley, A.~A., Morrison, H.~L., Helmi, A., et al.\ 2007, \aj, 134, 1579 
%\bibitem[Kinemuchi et al.(2006)]{kinemuchi2006} Kinemuchi, K., Smith, H.~A., Wo{\'z}niak, P.~R., McKay, T.~A., \& ROTSE Collaboration 2006, \aj, 132, 1202 
%\bibitem[Kinman et al.(2007)]{kinman2007} Kinman, T.~D., Cacciari, C., Bragaglia, A., Buzzoni, A., \& Spagna, A.\ 2007, \mnras, 375, 1381 
%\bibitem[Kinman et al.(2009)]{kinman2009} Kinman, T.~D., Morrison, H.~L., \& Brown, W.~R.\ 2009, \aj, 137, 3198 
%\bibitem[Koposov et al.(2010)]{koposov2010} Koposov, S.~E., Rix, H.-W., \& Hogg, D.~W.\ 2010, \apj, 712, 260 


%L

\bibitem[Layden(1998)]{layden1998} Layden, A.~C.\ 1998, \aj, 115, 
193 
\bibitem[Layden et al.(1999)]{layden1999} Layden, A.~C., Ritter, 
L.~A., Welch, D.~L., \& Webb, T.~M.~A.\ 1999, \aj, 117, 1313 


\bibitem[Law 
\& Majewski(2010)]{law2010} Law, D.~R., \& Majewski, S.~R.\ 2010, \apj, 718, 1128 
\bibitem[Layden et al.(1996)]{layden1996} Layden, A.~C., Hanson, R.~B., Hawley, S.~L., Klemola, A.~R., \& Hanley, C.~J.\ 1996, \aj, 112, 2110 


%M

\bibitem[Magnier et al.(2013)]{magnier2013} Magnier, E.~A., 
Schlafly, E., Finkbeiner, D., et al.\ 2013, \apjs, 205, 20 

\bibitem[Majewski et al.(2003)]{majewski2003} Majewski, S.~R., 
Skrutskie, M.~F., Weinberg, M.~D., 
\& Ostheimer, J.~C.\ 2003, \apj, 599, 1082 

\bibitem[McCarthy et al.(2012)]{mccarthy2012} McCarthy, I.~G., Font, 
A.~S., Crain, R.~A., et al.\ 2012, \mnras, 420, 2245 
\bibitem[Morganson et al.(2012)]{morganson2012} Morganson, E., De 
Rosa, G., Decarli, R., et al.\ 2012, \aj, 143, 142 
%\bibitem[Martin et al.(2004)]{martin2004} Martin, N.~F., Ibata, R.~A., Bellazzini, M., et al.\ 2004, \mnras, 348, 12 
%\\bibitem[Minniti et al.(2010)]{minniti2010} Minniti, D., Lucas, P.~W., Emerson, J.~P., et al.\ 2010, \na, 15, 433
%\\bibitem[Monet et al.(2003)]{monet2003} Monet, D.~G., Levine, S.~E., Canzian, B., et al.\ 2003, \aj, 125, 984 
%\\bibitem[Moody et al.(2003)]{moody2003} Moody, R., Schmidt, B., Alcock, C., et al.\ 2003, Earth Moon and Planets, 92, 125 
%\\bibitem[Morrison et al.(2009)]{morrison2009} Morrison, H.~L., Helmi, A., Sun, J., et al.\ 2009, \apj, 694, 130 

%N
%\bibitem[Newberg et al.(2002)]{newberg2002} Newberg, H.~J., Yanny,  B., Rockosi, C., et al.\ 2002, \apj, 569, 245 
\bibitem[Newberg et al.(2003)]{newberg2003} Newberg, H.~J., Yanny, 
B., Grebel, E.~K., et al.\ 2003, \apjl, 596, L191 
%\bibitem[Newberg et al.(2009)]{newberg2009} Newberg, H.~J., Yanny, B., \& Willett, B.~A.\ 2009, \apjl, 700, L61

%O

%P
\bibitem[Palaversa et al.(2013)]{palaversa2013} Palaversa, L., 
Ivezi{\'c}, {\v Z}., Eyer, L., et al.\ 2013, \aj, 146, 101 
%\bibitem[Pojmanski(2002)]{pojmanski2002} Pojmanski, G.\ 2002, \actaa, 52, 397 

%Q

%R
%\bibitem[Reimann(1994)]{reimann1994} Reimann, J.~D.\ 1994, Ph.D.~Thesis, Univ. California, Berkeley  

%S

\bibitem[Schlafly 
\& Finkbeiner(2011)]{schlafly2011} Schlafly, E.~F., \& Finkbeiner, D.~P.\ 2011, \apj, 737, 103 
\bibitem[Schlaufman et al.(2009)]{schlaufman2009} Schlaufman, K.~C., 
Rockosi, C.~M., Allende Prieto, C., et al.\ 2009, \apj, 703, 2177 
\bibitem[Schlaufman et al.(2012)]{schlaufman2012} Schlaufman, K.~C., 
Rockosi, C.~M., Lee, Y.~S., et al.\ 2012, \apj, 749, 77
\bibitem[Schlegel et al.(1998)]{schlegel1998} Schlegel, D.~J., 
Finkbeiner, D.~P., \& Davis, M.\ 1998, \apj, 500, 525 
%\bibitem[Schwarzenberg-Czerny(1989)]{stellingwerf1978} Schwarzenberg-Czerny, A.\ 1989, \mnras, 241, 153 
\bibitem[Sesar et al.(2007)]{sesar2007} Sesar, B., Ivezi{\'c}, 
{\v Z}., Lupton, R.~H., et al.\ 2007, \aj, 134, 2236 
\bibitem[Sesar et al.(2010)]{sesar2010} Sesar, B., Ivezi{\'c}, 
{\v Z}., Grammer, S.~H., et al.\ 2010, \apj, 708, 717 
\bibitem[Sesar et al.(2011)]{sesar2011} Sesar, B., Stuart, J.~S., 
Ivezi{\'c}, {\v Z}., et al.\ 2011, \aj, 142, 190 
\bibitem[Sesar et al.(2013)]{sesar2013} Sesar, B., Ivezi{\'c}, 
{\v Z}., Stuart, J.~S., et al.\ 2013, arXiv:1305.2160 


\bibitem[Smith(1995)]{smith1995} Smith, H.~A.\ 1995, RR Lyrae Stars. Cambridge Univ. Press, Cambridge
\bibitem[Stoughton et al.(2002)]{stoughton2002} Stoughton, C., 
Adelman, J., Annis, J.~T., et al.\ 2002, \procspie, 4836, 339 

%\bibitem[Skrutskie et al.(2006)]{skrutskie2006} Skrutskie, M.~F., Cutri, R.~M., Stiening, R., et al.\ 2006, \aj, 131, 1163
%\bibitem[Stellingwerf(1978)]{schwarzenberg1989} Stellingwerf, R.~F.\ 1978, \apj, 224, 953 
%\bibitem[Stokes et al.(2000)]{stokes2000} Stokes, G.~H., Evans, J.~B., Viggh, H.~E.~M., Shelly, F.~C., \& Pearce, E.~C.\ 2000, \icarus, 148, 21 

%T
\bibitem[Tonry et al.(2012)]{tonry2012} Tonry, J.~L., Stubbs, 
C.~W., Lykke, K.~R., et al.\ 2012, \apj, 750, 99 

%U
%\bibitem[Udalski et al.(2002)]{udalski2002} Udalski, A., Zebrun, K., Szymanski, M., et al.\ 2002, \actaa, 52, 115 
%\bibitem[Udalski(2003)]{udalski2003} Udalski, A.\ 2003, \actaa, 53, 291 

%V


\bibitem[VanderPlas et al.(2012)]{AstroML} VanderPlas, J., Connolly, A.~J., Ivezic, Z., \& Gray, A.\ 2012, Statistics, Data Mining and Machine Learning in Astronomy. Princeton Univ. Press, Princeton, NJ
%, pp.~47-54, 2012., 47 
%\bibitem[Vickers et al.(2012)]{vickers2012} Vickers, J.~J., Grebel, E.~K., \& Huxor, A.~P.\ 2012, \aj, 143, 86 
\bibitem[Vivas et al.(2001)]{vivas2001} Vivas, A.~K., Zinn, R., Andrews, P., et al.\ 2001, \apjl, 554, L33 
%\bibitem[Vivas et al.(2004)]{vivas2004} Vivas, A.~K., Zinn, R., Abad, C., et al.\ 2004, \aj, 127, 1158 
\bibitem[\protect\citeauthoryear{Vivas 
\& Zinn}{2006}]{vivas2006} Vivas A.~K., Zinn R., 2006, AJ, 132, 714 

%W
\bibitem[Watkins et al.(2009)]{watkins2009} Watkins, L.~L., Evans, 
N.~W., Belokurov, V., et al.\ 2009, \mnras, 398, 1757 
%\bibitem[Watkins et al.(2009)]{watkins2009} Watkins, L.~L., Evans, N.~W., Belokurov, V., et al.\ 2009, \mnras, 398, 1757 
%\bibitem[Welch \& Stetson(1993)]{welch1993} Welch, D.~L., \& Stetson, P.~B.\ 1993, \aj, 105, 1813 
%\bibitem[Wils et al.(2006)]{wils2006} Wils, P., Lloyd, C., \& Bernhard, K.\ 2006, \mnras, 368, 1757 

%X

%Y
%\bibitem[Yanny et al.(2000)]{yanny2000} Yanny, B., Newberg,  H.~J., Kent, S., et al.\ 2000, \apj, 540, 825 
\bibitem[Yanny et al.(2003)]{yanny2003} Yanny, B., Newberg, 
H.~J., Grebel, E.~K., et al.\ 2003, \apj, 588, 824 
\bibitem[York et al.(2000)]{york2000} York, D.~G., Adelman, J., 
Anderson, J.~E., Jr., et al.\ 2000, \aj, 120, 1579 

%Z
\bibitem[Zolotov et al.(2010)]{zolotov2010} Zolotov, A., Willman, 
B., Brooks, A.~M., et al.\ 2010, \apj, 721, 738 


\bibitem[Zinn et al.(2014)]{zinn2013} Zinn, R., Horowitz, B., 
Vivas, A.~K., et al.\ 2014, \apj, 781, 22 




\end{thebibliography}
\end{document}